\documentstyle[12pt]{article}

\font\twlgot =eufm10 scaled \magstep1
\font\egtgot =eufm8
\font\sevgot =eufm7

\font\twlmsb =msbm10 scaled \magstep1
\font\egtmsb =msbm8
\font\sevmsb =msbm7

\newfam\gotfam
\def\pgot{\fam\gotfam\twlgot}
\textfont\gotfam\twlgot
\scriptfont\gotfam\egtgot
\scriptscriptfont\gotfam\sevgot
\def\got{\protect\pgot}

\newfam\msbfam
\textfont\msbfam\twlmsb
\scriptfont\msbfam\egtmsb
\scriptscriptfont\msbfam\sevmsb

\def\pBbb{\relax\ifmmode\expandafter\Bb\else\typeout{You cann't use
Bbb in text mode}\fi}
\def\Bb #1{{\fam\msbfam\relax#1}}

\def\thebibliography#1{\bigskip\section*{\large
\bf References\\}\list
  {[\arabic{enumi}]}{\settowidth\labelwidth{#1}\leftmargin\labelwidth
    \advance\leftmargin\labelsep
    \usecounter{enumi}}
    \def\newblock{\hskip .11em plus .33em minus .07em}
    \sloppy\clubpenalty4000\widowpenalty4000
    \sfcode`\.=1000\relax}

\def\op#1{\mathop{{\it\fam0} #1}\limits}

\newcommand{\hm}{{\rm Hom\,}}
\newcommand{\dif}{{\rm Diff\,}}

\newcommand{\bll}{\bullet}
\newcommand{\beq}{\begin{equation}}
\newcommand{\eeq}{\end{equation}}
\newcommand{\ben}{\begin{eqnarray}}
\newcommand{\een}{\end{eqnarray}}
\newcommand{\be}{\begin{eqnarray*}}
\newcommand{\ee}{\end{eqnarray*}}
\newcommand{\bea}{\begin{eqalph}}
\newcommand{\eea}{\end{eqalph}}

\newcommand{\cO}{{\cal O}}
\newcommand{\cA}{{\cal A}}

\newcommand{\gd}{{\got d}}

\newcommand{\cI}{{\cal I}}

\newcommand{\cZ}{{\cal Z}}

\newcommand{\cK}{{\cal K}}

\newcommand{\dl}{\delta}

\newcommand{\f}{\phi}

\newcommand{\bb}{{\bf 1}}

\newcommand{\w}{\wedge}

\newcommand{\wh}{\widehat}
\newcommand{\ol}{\overline}

\newcommand{\dr}{\partial}

\newenvironment{eqalph}{\stepcounter{equation}
\setcounter{equationa}{\value{equation}}
\setcounter{equation}{0}

\begin{eqnarray}}{\end{eqnarray}\setcounter{equation}{\value{equationa}}}

\newcounter{example}
\newcounter{remark}
\newcounter{theorem}
\newcounter{proposition}
\newcounter{lemma}
\newcounter{corollary}
\newcounter{definition}

\def\theremark{\arabic{remark}}

\def\thedefinition{\arabic{definition}}

\newenvironment{proof}{\noindent{\it Proof.}}{\hfill $\Box$
\medskip }
\newenvironment{rem}{\refstepcounter{remark} \medskip \noindent{\bf Remark
\theremark.} \small}{ \medskip }

\newenvironment{prop}{\refstepcounter{definition} \medskip\noindent{\bf
Proposition \thedefinition.}}{\medskip }

\newenvironment{defi}{\refstepcounter{definition} \medskip\noindent{\bf 
Definition \thedefinition.} }{\medskip }

\newcommand{\mar}[1]{}

\begin{document}
\hbox{}

{\parindent=0pt

{\large \bf On the definition of higher order differential operators in 
noncommutative geometry}  
\bigskip 

{\bf G. Sardanashvily}

\medskip

\begin{small}

Department of Theoretical Physics, Moscow State University, 117234
Moscow, Russia

E-mail: sard@grav.phys.msu.su

URL: http://webcenter.ru/$\sim$sardan/
\bigskip

{\bf Abstract.}
Several definitions of differential operators on modules over 
a noncommutative ring are discussed. 
\end{small}
}

\bigskip
\bigskip

Let $\cK$ be a commutative ring and $\cA$ a $\cK$-ring (a unital
algebra with $\bb\neq 0$). Two-sided $\cA$-modules throughout are assumed 
to be (central) bimodules over the center $\cZ_\cA$ of $\cA$. 
Let $P$ and $Q$ be two-sided $\cA$-modules. We discuss the notion of a
 $\cK$-linear higher order $Q$-valued differential operator on $P$.
If a ring $\cA$ is commutative, there is the standard definition (up to
an equivalence) of a differential operator on $\cA$-modules \cite{grot,kras}. 
However, there exist its different generalizations to modules over a
noncommutative ring \cite{bor97,dublmp,lunts}. 
At the same time, derivations of a
noncommutative ring and the differential of a
differential calculus over a noncommutative ring
are defined in a standard way. It seems reasonable to regard them as
differential operators.

Let $\cA$ be a
commutative $\cK$-ring. Let $P$ and $Q$ be $\cA$-modules (central bimodules).
The $\cK$-module $\hm_\cK (P,Q)$
of $\cK$-linear homomorphisms $\Phi:P\to Q$ is endowed with the
two different $\cA$-module structures
\mar{5.29}\beq
(a\Phi)(p):= a\Phi(p),  \qquad  (\Phi\bll a)(p) := \Phi (a p),\qquad a\in
\cA, \quad p\in P. \label{5.29}
\eeq
We will refer to the second one as the $\cA^\bll$-module structure.
Let us put
\mar{spr172}\beq
\dl_a\Phi:= a\Phi -\Phi\bll a, \qquad a\in\cA. \label{spr172}
\eeq

\begin{defi} \label{ws131} \mar{ws131}
An element $\Delta\in\hm_\cK(P,Q)$ is an $r$-order $Q$-valued  
differential operator on $P$ if
$\dl_{a_0}\circ\cdots\circ\dl_{a_r}\Delta=0$
for any tuple of $r+1$ elements $a_0,\ldots,a_r$ of $\cA$.
\end{defi}

This definition is equivalent to the following one.

\begin{defi} \label{ws155} \mar{ws155}
An element $\Delta\in \hm_\cK (P,Q)$ is a zero order differential
operator if $\dl_a\Delta=0$ for all $a\in\cA$, and $\Delta$ is 
a differential operator of order $r>0$ if $\dl_a\Delta$ for all $a\in\cA$
is an $(r-1)$-order differential operator.
\end{defi}

In particular, zero order differential operators 
coincide with $\cA$-module 
morphisms $P\to Q$.
A first order differential operator
$\Delta$ satisfies the condition
\mar{ws106}\beq
\dl_b\circ\dl_a\,\Delta(p)= ba\Delta(p)-b\Delta(ap)
-a\Delta(bp) +\Delta(abp) =0, \qquad \forall a,b\in\cA.
\label{ws106}
\eeq

The set $\dif_r(P,Q)$ of $r$-order $Q$-valued 
 differential operators on $P$ inherits the $\cA$- 
and $\cA^\bll$-module structures (\ref{5.29}).

\begin{rem}
Let $P=\cA$. Any zero order $Q$-valued differential operator
$\Delta$ on $\cA$ is uniquely defined by its value $\Delta(\bb)$.
As a consequence, there is an isomorphism $\dif_0(\cA,Q)=Q$ 
via the association
$Q\ni q\mapsto \Delta_q$,
where $\Delta_q$ is given by the equality
$\Delta_q(\bb)=q$. 
A first order $Q$-valued differential
operator $\Delta$ on $\cA$ fulfils the condition
\be
\Delta(ab)=b\Delta(a)+ a\Delta(b) -ba \Delta(\bb), \qquad
\forall a,b\in\cA.
\ee
It is a $Q$-valued derivation of $\cA$ if
$\Delta(\bb)=0$, i.e., the
Leibniz rule
\mar{+a20}\beq
\Delta(ab) = \Delta(a)b + a\Delta(b), \qquad \forall a,b\in \cA, \label{+a20}
\eeq
holds. The set $\gd(\cA,Q)$ of derivations of $\cA$
is an $\cA$-module, but not an $\cA^\bll$-module.
Any first order differential operator on $\cA$
is split into the sum  
\be
\Delta(a)= a\Delta(\bb) +[\Delta(a)-a\Delta(\bb)]
\ee
of the zero order differential operator $a\Delta(\bb)$ and the derivation
$\Delta(a)-a\Delta(\bb)$.
As a consequence, there is the $\cA$-module decomposition 
\be
\dif_1(\cA,Q) = Q \oplus\gd(\cA,Q). 
\ee
\end{rem}

If $P$ and $Q$ are two-sided $\cA$-modules over a noncommutative 
$\cK$-ring $\cA$, the $\cK$-module $\hm_\cK(P,Q)$
can be provided with the left $\cA$- and right $\cA^\bll$-module structures
(\ref{5.29}) and the similar right and left structures
\mar{ws105}\beq
(\Phi a)(p):=\Phi(p)a, \qquad (a\bll\Phi)(p):=\Phi(pa), \qquad a\in\cA, \qquad
p\in\ P. \label{ws105}
\eeq
For the sake of convenience, we will refer to the $\cA-\cA^\bll$ structures
(\ref{5.29}) and (\ref{ws105}) as the left and right $\cA-\cA^\bll$
structures, respectively. 
Let us put 
\mar{ws133}\beq
\ol\dl_a\Phi:=\Phi a-a\bll\Phi, \qquad a\in\cA,
\qquad \Phi\in \hm_\cK(P,Q). \label{ws133}
\eeq
It is readily observed that  
$\dl_a\circ\ol\dl_b=\ol\dl_b\circ\dl_a$ for all $a,b\in\cA$.

The left $\cA$-module homomorphisms $\Delta: P\to Q$ obey the conditions
$\dl_a\Delta=0$, $\forall a\in\cA$, and, consequently, they can be
regarded as 
left zero order $Q$-valued differential operators on $P$. Similarly,
right zero order differential operators are defined.

Utilizing the
condition (\ref{ws106}) as a definition of a first order 
differential operator in noncommutative geometry, one however meets
difficulties. If $P=\cA$ and
$\Delta(\bb)=0$, the condition 
(\ref{ws106}) does not lead to the Leibniz rule (\ref{+a20}), i.e., derivations
of the $\cK$-ring $\cA$ are not first order differential operators.
In order to overcome these difficulties, one can replace the condition 
(\ref{ws106}) with the
following one \cite{dublmp}. 

\begin{defi} \label{ws120} \mar{ws120}
An element $\Delta\in \hm_\cK(P,Q)$ is called a first order
differential operator on a two-sided module $P$ over a noncommutative ring
$\cA$ if it obeys the relations
\mar{ws114}\ben
&& \dl_a\circ\ol\dl_b\Delta=\ol\dl_b\circ\dl_a\Delta=0,
\qquad \forall a,b\in\cA, \nonumber \\
&& a\Delta(p)b -a\Delta(pb) -\Delta(ap)b +\Delta(apb)=0, \qquad p\in P.
\label{ws114}
\een
\end{defi}

First order $Q$-valued differential operators on $P$ make up a
$\cZ_\cA$-module $\dif_1(P,Q)$.

If $P$ is a bimodule over a commutative ring $\cA$, then
$\dl_a=\ol\dl_a$ and Definition \ref{ws120} comes to Definition \ref{ws131}
for first order differential operators.

\begin{rem} Let $P=\cA$.
Any left or right zero order $Q$-valued differential operator $\Delta$
is uniquely defined by its value $\Delta(\bb)$. As a consequence, 
there are left and right  
$\cA$-module isomorphisms
\be
&& Q\ni q\mapsto \Delta^{\rm R}_q\in\dif_0^{\rm R}(\cA,Q), \qquad 
\Delta^{\rm R}_q(a)=qa, \qquad a\in\cA,\\
&& Q\ni q\mapsto \Delta^{\rm L}_q\in\dif_0^{\rm L}(\cA,Q), \qquad 
\Delta^{\rm L}_q(a)=aq.
\ee
A first order $Q$-valued differential operator $\Delta$ on $\cA$
fulfils the condition
\mar{ws110}\beq
\Delta(ab)=\Delta(a)b+a\Delta(b) -a\Delta(\bb)b. \label{ws110}
\eeq
It is a derivation of $\cA$ if $\Delta(\bb)=0$.
One obtains at once that any first order
differential operator on $\cA$ is split into the sums
\be
\Delta(a)=a\Delta(\bb) +[\Delta(a)-a\Delta(\bb)], \qquad
\Delta(a)=\Delta(\bb)a +[\Delta(a)-\Delta(\bb)a]
\ee
of the derivations $\Delta(a)-a\Delta(\bb)$ or
$\Delta(a)-\Delta(\bb)a$ and 
the left or right zero order differential operators $a\Delta(\bb)$ and 
$\Delta(\bb)a$, respectively. 
If $u$ is a $Q$-valued
derivation of $\cA$, then $au$ (\ref{5.29}) and $ua$ (\ref{ws105}) are
so for any $a\in\cZ_\cA$. Hence,
$Q$-valued derivations of $\cA$ constitute a $\cZ_\cA$-module $\gd(\cA,Q)$.
There are two
$\cZ_\cA$-module decompositions
\be
\dif_1(\cA,Q)= \dif_0^{\rm L}(\cA,Q) \oplus \gd(\cA,Q), \qquad
\dif_1(\cA,Q)= \dif_0^{\rm R}(\cA,Q) \oplus \gd(\cA,Q). 
\ee
They differ from each
other in the inner derivations $a\mapsto aq-qa$. 
\end{rem}

Let
$\hm_\cA^{\rm R}(P,Q)$ and 
$\hm_\cA^{\rm L}(P,Q)$ be the modules of right and left $\cA$-module 
homomorphisms of $P$ to $Q$, respectively. They are provided 
with the left and right $\cA-\cA^\bll$-module structures (\ref{5.29})
and (\ref{ws105}), respectively.

\begin{prop} \label{ws113} \mar{ws113}
An element $\Delta\in\hm_\cK(P,Q)$ is
a first order $Q$-valued differential operator on $P$ in accordance
with Definition \ref{ws120} iff it
obeys the condition
\mar{n21}\beq
\Delta(apb)=(\op\dr^\to a)(p)b +a\Delta(p)b + a(\op\dr^\leftarrow b)(p),
\qquad \forall p\in P, \qquad \forall a,b\in\cA,\label{n21}
\eeq
where $\op\dr^\to$ and $\op\dr^\leftarrow$ are 
$\hm_\cA^{\rm R}(P,Q)$- and 
$\hm_\cA^{\rm L}(P,Q)$-valued 
derivations of $\cA$, respectively. Namely, 
$(\op\dr^\to a)(pb)=(\op\dr^\to a)(p)b$ and $(\op\dr^\leftarrow b)(ap)
=a(\op\dr^\leftarrow b)(p)$. 
\end{prop}

\begin{proof}
It is easily verified that, if $\Delta$ obeys the equalities (\ref{n21}),
it also satisfies the 
equalities (\ref{ws114}). Conversely, let $\Delta$ be a first order $Q$-valued 
differential operator on $P$ in accordance
with Definition \ref{ws120}. One can bring the condition (\ref{ws114})
into the form
\be
\Delta(apb)=[\Delta(ap)-a\Delta(p)]b +a\Delta(p)b +a[\Delta(pb)-\Delta(p)b],
\ee
and introduce the derivations
\be
(\op\dr^\to a)(p):= \Delta(ap)-a\Delta(p), \qquad (\op\dr^\leftarrow b)(p):=
\Delta(pb)-\Delta(p)b.
\ee
\end{proof}

\begin{rem} \label{ws160} \mar{ws160}
Let $P$ be a differential calculus over a $\cK$-ring
$\cA$ provided with an associative multiplication $\circ$ and a
coboundary operator $d$. Then $d$ exemplifies a $P$-valued first order
differential operator on $P$ by Definition \ref{ws120}. It obeys the 
condition (\ref{n21}) which reads
\be
d(apb)=(da\circ p)b+a(dp)b + a((-1)^{|p|}p\circ db).
\ee
For instance, let $\gd\cA$ be the Lie $\cK$-algebra of $\cA$-valued
derivations of a $\cK$-ring $\cA$. Let us consider the
Chevalley--Eilenberg complex of the Lie $\cK$-algebra $\gd\cA$ with
coefficients in the ring 
$\cA$, regarded as a $\gd\cA$-module. This complex contains a
subcomplex $\cO^*[\gd\cA]$
of $\cZ_\cA$-multilinear skew-symmetric maps
$\f:\op\times^k \gd\cA\to \cA$
with respect to the Chevalley--Eilenberg coboundary
operator 
\be
&& d\f(u_0,\ldots,u_k)=\op\sum^k_{i=0}(-1)^iu_i
(\f(u_0,\ldots,\wh{u_i},\ldots,u_k)) +\\
&& \qquad \op\sum_{i<j} (-1)^{i+j}
\f([u_i,u_j],u_0,\ldots,
\wh u_i, \ldots, \wh u_j,\ldots,u_k). 
\ee
Its terms
$\cO^k[\gd\cA]$ are two-sided $\cA$-modules. In particular,
\mar{spr708,'}\ben
&& (d a)(u)=u(a), \qquad a\in\cA, \qquad u\in\gd\cA, \label{spr708}\\
&& \cO^1[\gd\cA]=\hm_{\cZ_\cA}(\gd\cA,\cA). \label{spr708'}
\een
The graded module $\cO^*[\gd\cA]$ is
provided with the product 
\be
\f\w\f'(u_1,...,u_{r+s})= 
\qquad \op\sum_{i_1<\cdots<i_r;j_1<\cdots<j_s}
{\rm sgn}^{i_1\cdots i_rj_1\cdots j_s}_{1\cdots r+s} \f(u_{i_1},\ldots,
u_{i_r}) \f'(u_{j_1},\ldots,u_{j_s}).
\ee
This product obeys the relation
\be
d(\f\w\f')=d(\f)\w\f' +(-1)^{|\f|}\f\w d(\f'),
\qquad \f,\f'\in \cO^*[\gd\cA],
\ee
and makes $\cO^*[\gd\cA]$ to a graded differential calculus
over $\cA$. This contains the graded differential subalgebra $\cO^*\cA$
generated by the elements $da$, $a\in\cA$. One can show that
\mar{ws130}\beq
\gd\cA=\hm_\cA(\cO^*\cA,\cA). \label{ws130}
\eeq
\end{rem}

In view of the relations (\ref{spr708'}) and (\ref{ws130}),
one can think of elements of $\cO^1\cA$ and $\gd\cA$ in Remark
\ref{ws160} as being 
differential one-forms and vector fields in noncommutative geometry.
A problem is that $\gd\cA$ is not an $\cA$-module. 
One can overcome this difficulty as follows \cite{bor97}. 

Given a noncommutative
$\cK$-ring $\cA$ and a two-sided $\cA$-module $Q$, let $d$ be a
$Q$-valued derivation of $\cA$. One can think of $Q$ as being a first
degree term of a differential calculus over $\cA$. Let $Q^*_{\rm R}$ be
the right $\cA$-dual of $Q$. It is a two-sided $\cA$-module:
\be
(bu)(q):= bu(q), \qquad (ub)(q):=u(bq), \qquad \forall b\in\cA, \qquad
q\in Q.
\ee
One can
associate to each element $u\in Q^*_{\rm R}$ the $\cK$-module morphism
\mar{ws121}\beq
\wh u:\cA\in a\mapsto u(da)\in\cA. \label{ws121}
\eeq
This morphism obeys the relations
\mar{ws123}\beq
\wh{(bu)}(a) =bu(da), \qquad \wh u(ba)=\wh u(b)a+\wh{(ub)}(a). \label{ws123}
\eeq
One calls $(Q^*_{\rm R},u\mapsto\wh u )$ the $\cA$-right Cartan pair,
and regards $\wh u$ (\ref{ws121}) as an $\cA$-valued first order differential
operator on $\cA$ \cite{bor97}. Note that $\wh u$ (\ref{ws121}) need not be a
derivation of $\cA$ 
and fails to satisfy Definition \ref{ws120}, unless $u$ belongs to the
two-sided $\cA$-dual 
$Q^*\subset Q^*_{\rm R}$ of $Q$. 

Morphisms $\wh u$
(\ref{ws121}) are called into play in order to describe (left) 
vector fields in
noncommutative geometry \cite{bor97,jara}. 

For instance, if $Q=\cO^1\cA$
in Remark \ref{ws160}, then $au$ for any $u\in\gd\cA$ and $a\in\cA$ is a left
noncommutative vector field in accordance with the relation (\ref{spr708}).

Similarly the $\cA$-left Cartan pair is defined. For instance,
$ua$ for any $u\in\gd\cA$ and $a\in\cA$ is a right
noncommutative vector field.

If $\cA$-valued derivation $u_1,\ldots u_r$ of a noncommutative
$\cK$-ring $\cA$ or the above mentioned noncommutative vector fields
$\wh u_1,\ldots \wh u_r$ on $\cA$ are regarded as first order
differential operators on $\cA$, it seems natural to think of their 
compositions $u_1\circ\cdots u_r$ or $\wh u_1\circ\cdots \wh u_r$ as
being particular higher order differential operators on $\cA$. 
Turn to the general
notion of a differential operator on two-sided $\cA$-modules.

Let $P$ and $Q$ be regarded as
left $\cA$-modules \cite{lunts}. 
Let us consider the $\cK$-module $\hm_\cK (P,Q)$ provided with the left
$\cA-\cA^\bll$ module structure (\ref{5.29}).
We denote $\cZ_0$ its center, i.e., $\dl_a\Phi=0$ for all $\Phi\in\cZ_0$
and $a\in\cA$. Let $\cI_0=\ol \cZ_0$ be the $\cA-\cA^\bll$ submodule 
of $\hm_\cK (P,Q)$ 
generated by $\cZ_0$. Let us consider: (i) the quotient $\hm_\cK (P,Q)/\cI_0$,
(ii) its center $\cZ_1$, (iii) the $\cA-\cA^\bll$ submodule $\ol \cZ_1$
of $\hm_\cK (P,Q)/\cI_0$ generated by $\cZ_1$, and (iv) the
$\cA-\cA^\bll$ submodule $\cI_1$  
of $\hm_\cK (P,Q)$ given by the relation 
$\cI_1/\cI_0=\ol \cZ_1$.
Then we
define the $\cA-\cA^\bll$ submodules $\cI_r$, $r=2,\ldots$, of
$\hm_\cK (P,Q)$ by induction
as follows:
\mar{ws134}\beq
\cI_r/\cI_{r-1}=\ol \cZ_r, \label{ws134}
\eeq
where $\ol \cZ_r$ is the $\cA-\cA^\bll$ module 
generated by the center $\cZ_r$ of the quotient $\hm_\cK (P,Q)/\cI_{r-1}$.
 
\begin{defi} \label{ws135} \mar{ws135}
Elements of the submodule $\cI_r$ of $\hm_\cK (P,Q)$ are said to be
left $r$-order $Q$-valued   
differential operators on a two-sided $\cA$-module $P$
\cite{lunts}. 
\end{defi}

\begin{prop} \label{ws137} \mar{ws137}
An element $\Delta\in \hm_\cK (P,Q)$ is a differential
operator of order $r$ in accordance with Definition \ref{ws135} iff
it is brought into a finite sum 
\mar{ws138}\beq
\Delta(p)=b_i\Phi^i(p) +\Delta_{r-1}(p), \qquad b_i\in\cA, \label{ws138}
\eeq
where $\Delta_{r-1}$ and $\dl_a\Phi^i$ for all $a\in\cA$ are left
$(r-1)$-order  
differential operators if $r>0$ and they vanish if $r=0$.
\end{prop}

\begin{proof}
If $r=0$, the statement is a straightforward consequence of 
Definition \ref{ws135}. Let $r>0$.
The representatives $\Phi_r$ of elements of $\cZ_r$ obey the relation
\mar{ws139}\beq
\dl_c\Phi_r= \Delta'_{r-1}, \qquad \forall c\in \cA, \label{ws139}
\eeq
where $\Delta'_{r-1}$ is an $(r-1)$-order differential operator. Then 
representatives $\ol\Phi_r$ of elements of $\ol\cZ_r$ take the form
\be
\ol\Phi_r(p)=\op\sum_i c'_i\Phi^i(c_ip) + \Delta''_{r-1}(p), \qquad
c_i,c'_i\in \cA, 
\ee
where $\Phi^i$ satisfy the relation (\ref{ws139}) and $\Delta''_{r-1}$
is an $(r-1)$-order differential operator. Due to the relation
(\ref{ws139}), we obtain
\mar{ws140}\beq
\ol\Phi_r(p)=b_i\Phi^i(p) + \Delta'''_{r-1}(p), \qquad b_i=c_ic'_i, \qquad
\Delta'''_{r-1}=-\op\sum_ic'_i\dl_{c_i}\Phi^i + \Delta''_{r-1}. \label{ws140}
\eeq
Hence, elements of $\cI_r$ modulo elements of $\cI_{r-1}$ take the form
(\ref{ws140}), i.e., they are given by the expression (\ref{ws138}).
The converse is obvious.
\end{proof}

If $\cA$ is a commutative ring,
Definition \ref{ws135} comes to Definition \ref{ws155}. 
Indeed, the expression
(\ref{ws138}) shows that
$\Delta\in \hm_\cK (P,Q)$ is an $r$-order
differential operator iff $\dl_a\Delta$ for all $a\in\cA$ is a
differential operator of order $r-1$.

\begin{prop} \label{ws143} \mar{ws143}
The set $\cI_r$ of
$r$-order $Q$-valued differential operators on $P$ is provided with the
left and right $\cA-\cA^\bll$ module structures.
\end{prop}

\begin{proof}
This statement is obviously true for zero order differential operators.
Using the expression
(\ref{ws138}), one can prove it for higher order 
differential operators by
induction.
\end{proof}

Let $P=Q=\cA$. Any zero order
differential operator on $\cA$ in accordance with Definition \ref{ws135}
takes the form
\be
a\mapsto \op\sum_ic_iac'_i, \qquad c_i,c'_i\in\cA.
\ee

\begin{prop} \label{ws146} \mar{ws146}
Let $\Delta_1$ and $\Delta_2$ be $n$- and $m$-order $\cA$-valued 
differential
operators on $\cA$, respectively. Then their composition
$\Delta_1\circ\Delta_2$ is an 
$(n+m)$-order differential operator.
\end{prop}

\begin{proof}
The statement is proved by induction as follows.
If $n=0$ or $m=0$, it issues from the fact that 
the set of differential operators possesses both left and right 
$\cA-\cA^\bll$ structures. Let us assume that 
$\Delta\circ\Delta'$ is a differential operator for any $k$-order differential
operators $\Delta$ and $s$-order differential operators $\Delta'$ such
that $k+s<n+m$. Let us show that 
$\Delta_1\circ\Delta_2$ is a  differential operator of order $n+m$. Due to
the expression (\ref{ws138}), it
suffices to prove this fact when $\dl_a\Delta_1$ and $\dl_a\Delta_2$
for any $a\in\cA$ are differential operators of order $n-1$ and $m-1$,
respectively. We have the equality
\be
&& \dl_a(\Delta_1\circ\Delta_2)(b)=
a(\Delta_1\circ\Delta_2)(b)-(\Delta_1\circ\Delta_2)(ab)= \\
&& \qquad 
\Delta_1(a\Delta_2(b))+(\dl_a\Delta_1\circ\Delta_2)(b)-
(\Delta_1\circ\Delta_2)(ab) =  (\Delta_1\circ\dl_a\Delta_2)(b)+
(\dl_a\Delta_1\circ\Delta_2)(b),
\ee
whose right-hand side, by assumption, is a differential operator 
of order $n+m-1$.
\end{proof}

Any derivation $u\in\gd\cA$ of a $\cK$-ring $\cA$ is a first order
differential 
operator in accordance with Definition \ref{ws135}. Indeed, it is
readily observed that
\be
(\dl_au)(b)= au(b)-u(ab)=-u(a)b, \qquad b\in\cA,
\ee
is a zero order differential operator for all $a\in\cA$. 
The compositions $au$, $u\bll a$
(\ref{5.29}), $ua$, 
$a\bll u$ (\ref{ws105}) for any $u\in\gd\cA$, $a\in\cA$ and the
compositions of derivations  
$u_1\circ\cdots\circ u_r$ are also differential operators on $\cA$ 
in accordance with Definition
\ref{ws135}. 

At the same time, noncommutative vector fields 
do not satisfy Definition \ref{ws135} in general.  
First order differential
operators by Definition \ref{ws120} also need not obey Definition
\ref{ws135}, unless $P=Q=\cA$.

By analogy with Definition \ref{ws135} and Proposition \ref{ws137}, one
can define right differential operators on a two-sided $\cA$-module $P$
as follows.

\begin{defi} \label{ws151} \mar{ws151}
Let $P$ and $Q$ be seen as right $\cA$-modules over a noncommutative
$\cK$-ring $\cA$. 
An element $\Delta\in\hm_\cK(P,Q)$ is said to be a right zero order
$Q$-valued differential operator on $P$ if it is a finite sum
$\Delta=\Phi^i b_i$, $b_i\in\cA$, 
where $\ol\dl_a\Phi^i=0$ for all $a\in\cA$.
An element $\Delta\in\hm_\cK(P,Q)$
is called a right
differential operator of order $r>0$ on $P$ if it is a finite sum
\mar{ws150}\beq
\Delta(p)=\Phi^i(p)b_i +\Delta_{r-1}(p), \qquad b_i\in\cA, \label{ws150}
\eeq
where $\Delta_{r-1}$ and $\ol\dl_a\Phi^i$ for all $a\in\cA$ are right 
$(r-1)$-order differential operators.
\end{defi}

Definition \ref{ws135} and Definition
\ref{ws151} of left and right differential operators on two-sided
$\cA$-modules are 
not equivalent, but one can combine them as follows. 

\begin{defi} \label{ws152} \mar{ws152}
Let $P$ and $Q$ be two-sided modules over a noncommutative $\cK$-ring $\cA$.
An element $\Delta\in\hm_\cK(P,Q)$ is a two-sided zero order $Q$-valued 
differential operator on $P$ if 
it is either a left or right zero order differential operator.
An element $\Delta\in\hm_\cK(P,Q)$ is said to be a two-sided differential
operator of order $r>0$ on $P$ if 
it is brought both into the form $\Delta=b_i\Phi^i +\Delta_{r-1}$, $b_i\in\cA$,
and 
$\Delta=\ol\Phi^i\ol b_i +\ol\Delta_{r-1}$, $\ol b_i\in\cA$,
where $\Delta_{r-1}$, $\ol\Delta_{r-1}$ and
$\dl_a\Phi^i$, $\ol\dl_a\ol\Phi^i$ for all $a\in\cA$  
 are two-sided $(r-1)$-order differential operators.
\end{defi}

One can think of this definition as a generalization of
Definition \ref{ws120} to higher order differential operators.

It is readily observed that two-sided differential operators described by
Definition \ref{ws152} need not be left or right differential
operators, and {\it vice versa}.   
At the same time, $\cA$-valued derivations of a $\cK$-ring $\cA$ and their
compositions obey Definition \ref{ws152}.

\end{document}